\documentclass[aps,pra,reprint,10pt,superscriptaddress,amssymb,amsfonts,longbibliography,nofootinbib]{revtex4-1}

\usepackage{amsmath}
\usepackage{graphicx}
\usepackage{dcolumn}
\usepackage{physics}
\usepackage{bbm}
\usepackage{subfigure}
\usepackage{amssymb}
\usepackage{wasysym}
\usepackage{MnSymbol}
\usepackage{color}
\usepackage{mathrsfs}
\usepackage[normalem]{ulem}
\usepackage{titlesec}
\usepackage{soul}

\stepcounter{secnumdepth}
\stepcounter{tocdepth}

\newcommand{\btext}[1]{{\color{black}#1}}
\newcommand{\rtextnpj}[1]{{\color{black}#1}}

\begin{document}

\title{Fractons from frustration in hole-doped antiferromagnets}

\author{John Sous}\thanks{js5530@columbia.edu}
\affiliation{Department of Physics, Columbia University, New York, New York 10027, USA} 

\author{Michael Pretko}\thanks{michael.pretko@colorado.edu}
\affiliation{Department of Physics and Center for Theory of Quantum Matter, University of Colorado, Boulder, CO 80309, USA}

\date{\today}

\begin{abstract} 

Recent theoretical research on tensor gauge theories led to the discovery of an exotic type of quasiparticles, dubbed fractons, that obey both charge and dipole conservation.  Here we describe physical implementation of dipole conservation laws in realistic systems.  We show that fractons find a natural realization in hole-doped antiferromagnets. There, individual holes are largely immobile, while dipolar hole pairs move with ease.  First, we demonstrate a broad parametric regime of fracton behavior  in hole-doped two-dimensional Ising antiferromagnets viable through five orders in perturbation theory.  We then specialize to the case of holes confined to one dimension in an otherwise two-dimensional antiferromagnetic background, which can be realized via the application of external fields in experiments, and prove ideal fracton behavior. We explicitly map the model onto a fracton Hamiltonian featuring conservation of dipole moment.  Manifestations of fractonicity in these systems include gravitational clustering of holes.  We also discuss diagnostics of fracton behavior, which we argue is borne out in existing experimental results.
\end{abstract}

\titleformat{\section}{\normalfont\fontsize{12}{15}\raggedright\bfseries}{\arabic{section}.}{10em}{}
\titleformat{\subsection}{\raggedright\bfseries}{\arabic{section}.}{1em}{}
\titleformat{\subsubsection}{\raggedright\bfseries\it}{\arabic{section}.}{1em}{}

\titlespacing\subsection{0pt}{12pt plus 4pt minus 2pt}{0pt plus 2pt minus 2pt}
\titlespacing\subsubsection{0pt}{12pt plus 4pt minus 2pt}{0pt plus 2pt minus 2pt}

\maketitle

\normalsize

 \section*{Introduction}
 
The concept of exotic emergent quasiparticles has played a prominent role in the theory of strongly correlated quantum many-body systems for several decades, appearing in contexts ranging from fractional quantum Hall systems \cite{Hansson} to quantum spin liquids \cite{savary}.  Recently, an exotic type of emergent quasiparticle has been proposed: Fracton particles that exhibit an unusual form of mobility.  An individual fracton is strictly locked in place, while bound states of paired fractons are free to move around the system \cite{chamon,haah,fracton1,fracton2,sub}.  Fractons have drawn immense excitement partly because of their promise as a potential platform for fault-tolerant quantum computation and robust quantum information storage \cite{haah,bravyi,terhal}. But even more, their fundamental features are interesting in their own right, leading to  deep connections with a wide variety of concepts, such as tensor gauge theories \cite{sub,higgs1,higgs2}, gravity \cite{mach,holo1}, and localization \cite{chamon,glassy,spread,spread2,pollmann,KhemaniScars,confine}.  We refer the reader to a review article \cite{review} and selected literature \cite{
hanlayer,field,generic,phases,albert,deconfined,fractalsym,potter,supersolid,cheng} for further details.

 \begin{figure}[!htb]
\begin{flushleft}
\hspace{0cm} {\bf a)}
\end{flushleft}
\vspace{-0.3cm}
\includegraphics[width=\columnwidth]{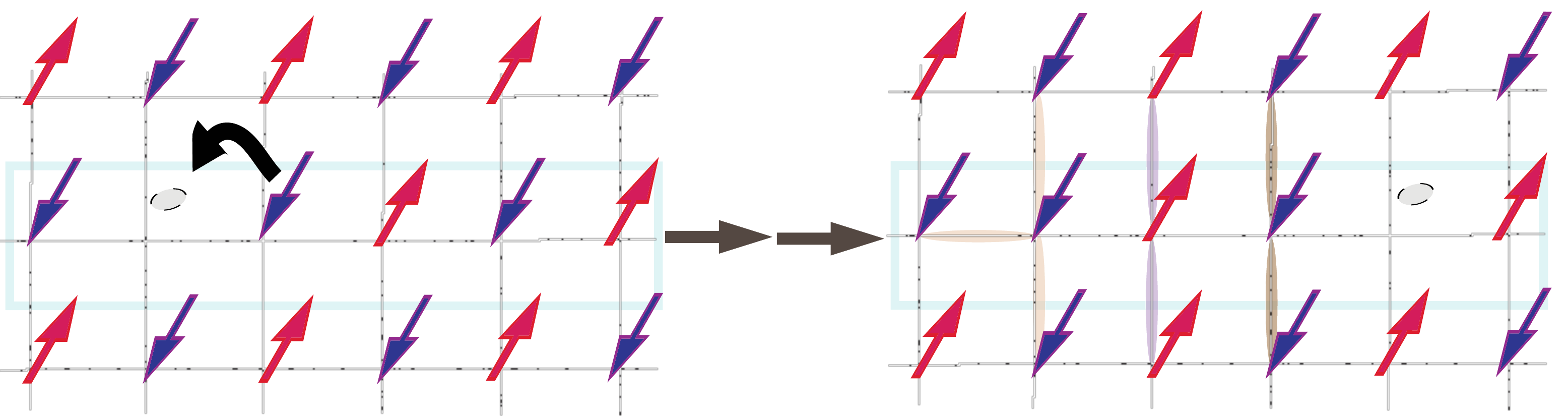} 
\begin{flushleft}
\hspace{0cm} {\bf b)}
\end{flushleft}
\vspace{-0.3cm}
  \includegraphics[width=\columnwidth]{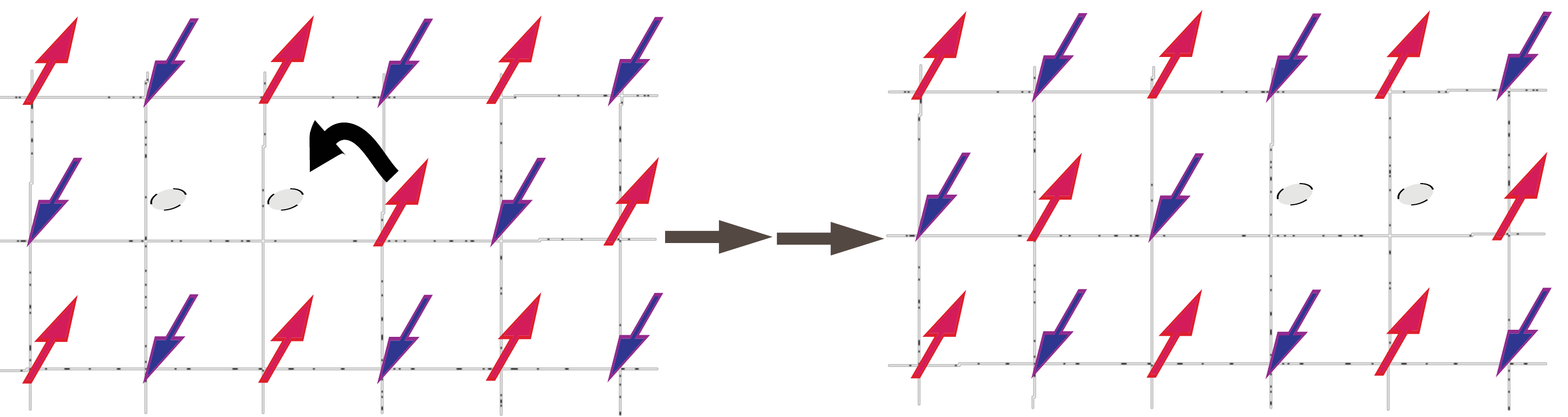}
 \caption{{\bf Fractons in hole-doped antiferromagnets.} {\bf a)}:  Motion of a single hole in a  (mixed-dimensional) N\'eel antiferromagnet frustrates the antiferromagnetic bonds.  {\bf b)}:  Two holes in a dipolar bound state move without breaking any antiferromagnetic bonds.
 }
 \label{fig:Holes}
 \end{figure}

The unusual fracton mobility constraints can be conveniently encoded as charge and dipole conservation laws.  While nature readily supplements charge symmetry in a plethora of physical systems, the realization of dipolar symmetry in realistic systems represents a challenge.  The ramifications on physical behavior of this aberrant symmetry constraint partially accounts for the widespread research activity on fractons.  Yet only few proposals for their realization in concrete physical systems exist.  One promising direction proposed fractons realized as disclination defects of two-dimensional crystals, with striking manifestations, such as glide constraint on dislocations \cite{elasticity,gromov,pai}. Unfortunately, the study of individual fractons in these systems is unfeasible due to the large energy cost required to separate disclinations.  Another significant push towards making contact with experiment involves engineering realistic fracton spin-liquid models \cite{slaglenn,rank2ice}.  However, as of now, no specific material candidates for a fracton spin liquid exist.  It may also be possible to impose closely related conservation laws in engineered cold-atom systems via application of a linear potential \cite{refael,huse}.  Emergence of fracton physics in these systems, however, remain to be seen.  It is therefore of paramount importance to identify realistic platforms for fracton physics, where individual fractons can be probed and analyzed, permitting a controllable study of few- to many-body behavior of fractonic systems.

In this paper, we identify one such platform, and explain that hole-doped antiferromagnets (AFMs) realize fracton physics at the single-, few- and many-particle levels. While the mobility restrictions of fractons may seem exotic at first glance, strikingly similar phenomenology is found in the simple, familiar physical setting of holes doped into an Ising antiferromagnet in dimensions greater than one: Motion of a single hole through the antiferromagnetic background is inhibited by creation of magnon (spin-flip) excitations \cite{Trugmanloops,ShraimanHole,KaneHole,SachdevHole,ChernHole}, see Figure \ref{fig:Holes}, {\bf a)}. Meanwhile, a bound pair of holes can easily move through the system, in a manifestation of fracton physics, as shown in Figure \ref{fig:Holes}, {\bf b)}.  In a fully two-dimensional (2D) AFM, this fracton behavior is only approximate due to higher-order Trugman loops that induce mobility of a single hole \cite{Trugmanloops}. However, since holes move only at sixth order, while dipolar bound states move at second order, the system features a wide parametric regime of fracton behavior.

While the 2D AFM exhibits only approximate fracton behavior, we next investigate a sharp realization of fracton behavior, specializing to the case of holes confined to one dimension of an otherwise two-dimensional antiferromagnetic background, a setup that can be achieved in experiments \cite{demler}.  In this system, we show that Trugman loops are entirely eliminated and the system exhibits perfect fracton behavior to all orders.  By integrating out the magnons, we explicitly derive an effective fracton Hamiltonian for the holes, characterized by conservation of dipole moment.

The manifestation of fracton physics in these systems, whether exact or approximate, has important consequences, some of which have already been borne out in existing experimental results.  Most notably, fractons exhibit a universal short-ranged `gravitational' attraction that can cause them to cluster together \cite{mach}.  This gravitational attraction coincides with the magnon-mediated interaction between holes, which has been identified as a potential pairing `glue' in superconductivity \cite{Exp1, glue}. A finite density of holes doped into an antiferromagnet experience phase separation \cite{emery,kivelson,Marder,batista}, in agreement with clustering and emulsion physics encountered in fracton theories \cite{phases}. To elucidate the fracton nature of the underlying excitations in these systems, we identify a signature in the pair correlation function as a diagnostic of fracton behavior.  Our results and calculations can be appropriately extended to more general classes of systems described by boson-affected hopping.  

\section*{Results}

\subsection*{Fractons in hole-doped antiferromagnets} 

\noindent{\em Two-dimensional square antiferromagnets:} We consider a small number of holes doped into a two-dimensional (2D) square Ising antiferromagnet (AFM) described by the Hamiltonian $H_{\rm Ising} = J \sum_{\langle i,j\rangle } S^z_i S^z_j$, $J>0$.  The undoped parent ground state of such a system is a classical N\'eel state: $\ket{\Psi_{\rm GS}} = \Pi_{i\in A} c^\dagger_{i,\uparrow} \Pi_{j\in B} c^\dagger_{j,\downarrow}\ket{0}$ with spins on sublattice $A$ pointing up, and spins on sublattice $B$ pointing down. Here, a $c^\dagger_{\sigma}$ operator creates a fermion with spin $\sigma = \{\uparrow, \downarrow \}$. A doped hole moves through motion of a spin particle to the empty site.  To very good approximation, holes move only via nearest-neighbor hopping.

The motion of the hole occurs as a result of the hopping of a particle whose spin becomes either perfectly aligned or misaligned with respect to the antiferromagnetic environment.  We can regard a misaligned spin as a bosonic defect, i.e. a magnon, with a creation operator:
\begin{eqnarray}
d_i^\dagger = \begin{cases}
     \sigma_i^-, & \text{if $i \in A$},\\
     \sigma_i^+, & \text{if $i \in B$},
  \end{cases}
\end{eqnarray}
where $\sigma^\pm$ is the spin-$\frac{1}{2}$ raising/lowering Pauli matrix. It is important to note that the hole motion conserves the total magnetization (as well as the charge) of the doped system.  One can therefore associate the removal of a fermion of spin $\sigma$ with the creation of a hole with spin $-\sigma$, as either amounts to a total net change of the magnetization of the entire system by $-\sigma$. We thus define the hole operators as
\begin{eqnarray}
h_{i,-\sigma}^\dagger =      c_{i,\sigma};
\sigma =\begin{cases}
     \uparrow , & \text{if $i \in A$},\\
     \downarrow, & \text{if $i \in B$}.
  \end{cases}
\end{eqnarray}
When the hole moves it either creates a magnon at the site of its departure (displaced oppositely oriented spin) or absorbs a magnon at the site of its arrival (heals a spin misalignment).  This gives a Hamiltonian \cite{MonaFehske2D}:
\begin{eqnarray}
H = \mathcal{P}\Bigg[-t \sum_{\langle i,j \rangle, \sigma}  \left[ h^\dagger_{j,\sigma} h_{i,\sigma} (d_i^\dagger + d_j) + h.c.\right] \Bigg] \mathcal{P}+ H_{{\rm Ising}}. \nonumber \\
\end{eqnarray}
Here $\langle . \rangle$ refers to nearest neighbors, and the Hamiltonian respects a no-double occupancy constraint implemented via a projector $\mathcal{P}$, so that each site has either a hole or a spin: $\sum_{\sigma}h_{i,\sigma}^\dagger h_{i,\sigma} + d_i^\dagger d_i + d_i d_i^\dagger=1$.  

Consider first a single hole doped into the 2D AFM.  The $\sigma$ label for the hole flavor is irrelevant and simply drops. The hole moves through the hopping of a fermion to the hole's original site. One must ask how the coupling of the hole hopping to magnons affects its motion. To address this question, we expand about the limit of a static hole in orders of $t/J$. The leading-order process is one in which the hole hops creating a misaligned spin or a magnon at the site of its departure, and then hops back to absorb the magnon, healing the background.  Continuing with a detailed analysis of this perturbative expansion reveals that the hole creates a string of spin flips as it moves, only to retrace them back to its original site, i.e. the hole is localized to its original site by the energetically costly strings.  Deviations from this picture occur at an order sixth in the expansion, corresponding to the motion of the hole in closed loops, known as Trugman loops \cite{Trugmanloops}, in which case it heals the string terminating two sites apart from its original site.  This analysis asserts that a single hole is localized through five orders in perturbation theory.

Consider now two holes of different spin flavor doped into the 2D AFM.  Holes exchange magnons, and thus interact. As before, we study the behavior via a perturbative expansion in $t/J$. To leading order, one hole moves and creates a magnon at its departure site, which is absorbed either by the second hole of opposite flavor, mediating the motion of the two-hole state or by the first hole, restoring the original configuration.  Since a hole cannot be simultaneously at the same site as a magnon, holes communicate only via strings.  Thus, to arbitrary order, we conclude that two holes are bound by a string and their motion is described by an effective pair-hopping interaction that moves the pair as a whole whilst preserving their relative distance.

We ascribe an effective charge degree of freedom to the magnetic polaron's spin: \rtextnpj{$\rho= h^\dagger \sigma^z h$, where $\sigma^z$ is the Pauli matrix in the spin flavor subspace of the fermion with eigenvalues $\sigma=\pm1$}. A single `charge', i.e. a polaron, cannot move in isolation, through five orders in perturbation theory. Two opposite charges, i.e. within a bipolaron, move together preserving their relative separation, and whence the bound state dipole moment: \rtextnpj{$D = \sum_i (h_i^\dagger \sigma^z_i h_i) x_i$}. This theory manifestly gives rise to a dipole conservation law, $\sum_i\,\rho_i x_i = \textrm{constant}$, i.e. a parametric regime of fractonic behavior, only violated at the sixth order in perturbation theory when a single hole becomes mobile. 

\noindent{\em Mixed-dimensional antiferromagnets:} One can achieve ideal fracton behavior in an intermediate setup between one and two dimensions, in so-called `mixed-dimensional' AFMs.  Applying a strong gradient potential $V(y)$ along the $y$-direction (taken to be one of the principal axes of the square lattice) restricts the hole to a line along the $x$-direction \cite{demler}.  This eliminates the undesirable motion of the hole along closed loops, while preserving spin frustration induced by hole motion, the mechanism behind string-mediated localization of the hole.

In this mixed dimensionality limit, a single hole always creates magnons first before absorbing them in the {\it reverse} order.  This simplifies the equation of motion for the one-hole propagator $G_{1h}(k,\omega)$, see {Methods} and Supplementary Note 1.  We find  for the lowest pole $G_{1h}(k,\omega) \sim [\omega - E_{\rm p}(k)]^{-1}$, where $E_{\rm p}(k)$ is the energy dispersion of a magnetic polaron formed as the background fluctuations dress the hole. We find $E_{\rm p}(k)$ to be dispersionless, reflecting the localization of the hole at its original position by string excitations.  Insight into this process can be gained as follows.  To leading (second) order, the polaron energy is $E_{\rm p}(k) = -4t^2/3J$, reflecting a process in which the hole hops from site $i_x$ to $i_x\pm1$ via one application of the hoping operator with amplitude $t$, creating a magnon with energy $3J/2$ at $i_x\pm1$, which it then absorbs and moves back to $i_x$. The extra factor of $2$ accounts for the two possible directions of hops in the $x$-direction. The next correction goes as $\sim t^4/J^3$.  Importantly, there is no possible way for the hole to end up at a site different from its original one, reflecting the fact that the hole always retraces its path back to the origin. 

Consider now the two-hole propagator $G_{2h}(K,\omega)$ in the $\sigma^z_{\text{Total}} = 0$ sector.  Since one hole of spin $\sigma$ always first emits a string of magnons before they are absorbed by the second hole of spin $-\sigma$, this propagator is also computed exactly self-consistently, see {Supplementary Note 1}.  This interaction binds the two magnetic polarons into a bipolaron via a string, with a dispersion: $E_{\rm BP} (K) = -\sum_{\delta=1}2(_2t_{\delta})\cos(\delta K)$, where $_2t_{\delta}$ is the amplitude of hopping of the composite bipolaron with momentum $K=k_{\sigma}+k_{-\sigma}$ $\delta$ sites and the discrete sum over $\delta$ truncates at some order in the expansion.  To leading order, the bipolaron dispersion is $-2(_2t_1)\cos(K)$ and $_2t_1 = 4t^2/3J$.  This dispersion reflects a magnon-mediated pair-hopping interaction that moves a pair of holes as a whole: $h_{(i_x,i_y)}^\dagger h_{(i_x+1,i_y)}^\dagger \ket{0} \rightarrow h_{(i_x+1,i_y)}^\dagger h_{(i_x+2,i_y)}^\dagger \ket{0}$/ $h_{(i_x-1,i_y)}^\dagger h_{(i_x,i_y)}^\dagger \ket{0}$; here $\ket{0} \equiv \ket{\Psi_{\rm GS}}$ is the vacuum of holes.  Note that the relative distance between the holes in the bipolaron always remain conserved.

We can calculate the two-particle behavior to arbitrary order, finding the effective Hamiltonian governing the one- and two-hole physics, after integrating out the bosons, to be:
\begin{eqnarray}\label{FracH}
 &&\quad H = -\epsilon_0 \sum_{i,\sigma} h^\dagger_{i,\sigma} h_{i,\sigma} \nonumber \\
&&- \sum_{i,\delta,\sigma} t_\delta \left(h_{i+\delta,\sigma}^\dagger h_{i+\delta+1,-\sigma}^\dagger + h_{i-\delta,\sigma}^\dagger h_{i-\delta+1,-\sigma}^\dagger\right)  h_{i+1,-\sigma} h_{i,\sigma} + \cdot \cdot \cdot, \nonumber \\
\end{eqnarray}
where $\epsilon_0$ is the polaron formation energy (discussed above) that gives rise to a simple shift in the particle's energy, and the $\cdot \cdot \cdot$ refers to other two-body density-density interactions. Here $i$ refers to the site index along a line in $x$. Importantly, this Hamiltonian does not generate any single-particle motion, but induces two-particle dynamics that preserves the relative distance between the two different polaron flavors, i.e. the two fractons form a composite particle (dipole) with a fixed radius.  As such, the dipole moment is strictly conserved, $[H,\rho]=[H,D] = 0$, representing perfect fractonicity.

\subsection*{Manifestations of fracton behavior}
\noindent A hallmark of fracton behavior is the presence of a universal attraction between fractons that can be regarded as an emergent gravitational force \cite{mach}, which we show leaves its signatures in hole-doped antiferromagnets.  This attraction arises as a consequence of the fact that fractons are more mobile in the vicinity of other ones.  Consider a particle with an effective mass $m(r)$, where $r$ is the distance away from a second particle in the system, taken to be fixed.  Neglecting inter-particle interactions, the fracton's velocity is $v = \sqrt{\frac{2E}{m(r)}}$: The velocity of the particle increases at small inter-particle separation and decreases otherwise. This attraction holds for both perfect and approximate fracton behavior, so long as $m(r)$ increases as particles move apart.

This effective attraction also continues to hold even in the presence of a sufficiently weak short-range repulsion $V(r) = V_0 e^{-r/a}$ between the holes, where $a$ is the lattice scale \btext{($V_0$ is of order $t^2/J$, corresponding to one of the ``$...$'' terms of Equation \ref{FracH})}.  Then, the velocity of a particle becomes $v = \sqrt{\frac{2(E-V(r))}{m(r)}}$.  Let us consider a generic case for the behavior of $m(r)$ with distance: \btext{We take its decay at short distances to be short-ranged, $i.e.$ $m(r) = m_0(1-\eta e^{-r/a})$ with $\eta < 1$ sets the energy scale of the dynamics of the fracton.} (Note that we let $V(r)$ and $m(r)$ range similarly, which is a useful simplifying assumption, though not fundamental to the analysis.)  \btext{Microscopically, the gravitational mass corresponds to the inverse of the hopping. Since the hole hopping (in 2D AFMs) $\sim t^6/J^5$, while the partner-induced hopping $\sim t^2/J$, we can extract $m_0\sim J^5/t^6 $ and $\eta/m_0  \sim t^2/J \rightarrow \eta \sim (J/t)^4$.}  The velocity of a particle is then:
\begin{equation}
v = \sqrt{\frac{2(E-V_0e^{-r/a})}{m_0(1-\eta e^{-r/a})}}\approx \sqrt{\frac{2E}{m_0}}\bigg(1+\frac{1}{2}(\eta - \frac{V_0}{E})e^{-r/a}\bigg).
\end{equation}
The last step represents the leading behavior at large $r$, i.e. bigger than a few lattice spacings.  As long as $V_0$ remains sufficiently weak, such that $V_0 < \eta E$, the effective force between fractons will remain attractive for the majority of states.  \btext{Since $\eta\sim (J/t)^4$ while $V_0\sim t^2/J$, this condition will hold for nearly all states.}  (This condition will fail for certain configurations with sufficiently small $E$.  However, such states typically involve widely separated particles that do not interact significantly  anyway.)  We therefore see that particles governed by approximate fracton behavior will still exhibit near-universal attraction, even in the presence of a small short-range repulsion.

\begin{figure}[!tb]
\begin{flushleft}
\hspace{0cm} {\bf a)} \hspace{4.2cm} {\bf b)}
\end{flushleft}
  \hspace{-0.3cm} \includegraphics[scale=0.1, height=0.35\columnwidth]{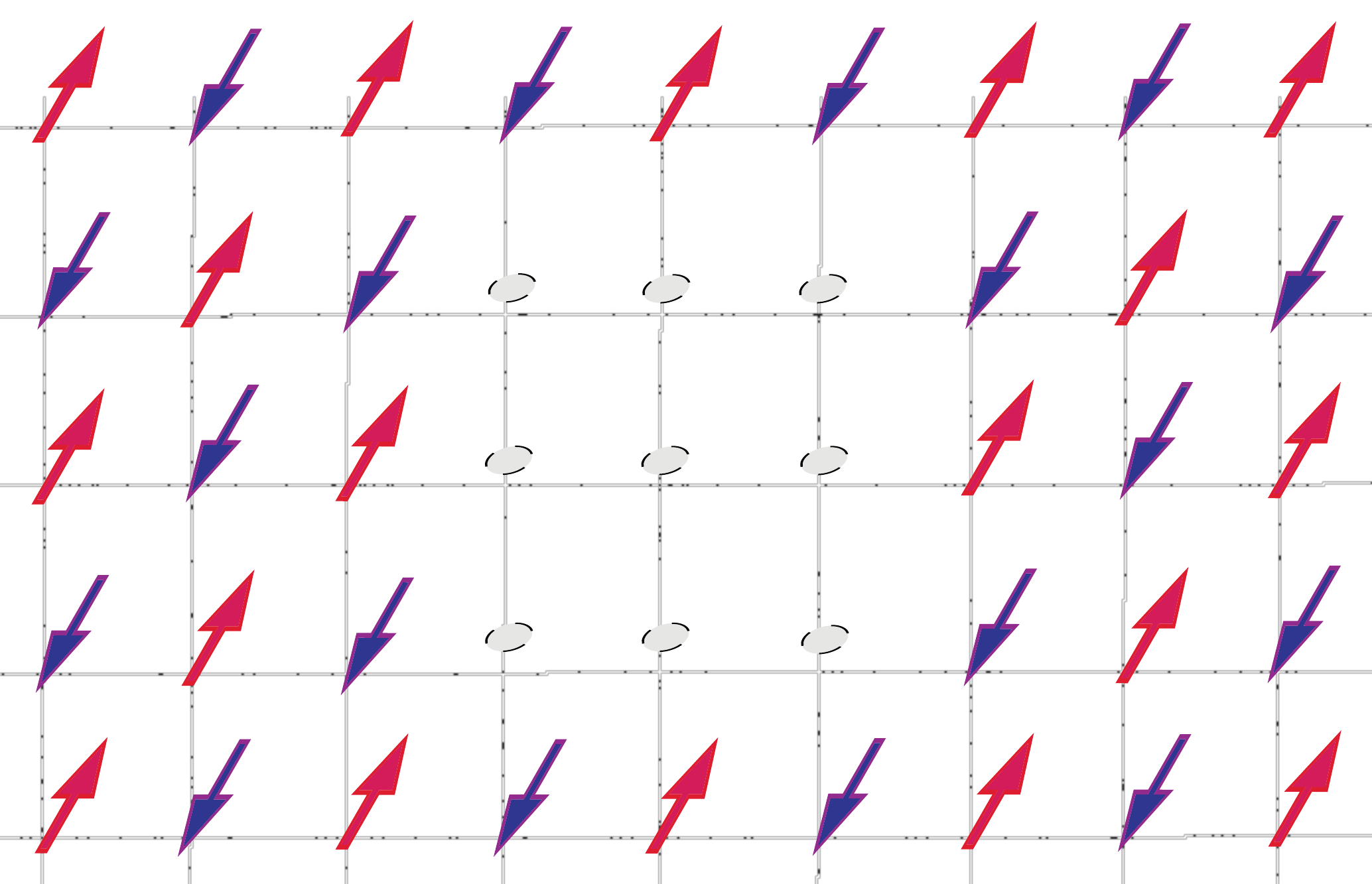} 
        \includegraphics[scale=0.1, height=0.325\columnwidth]{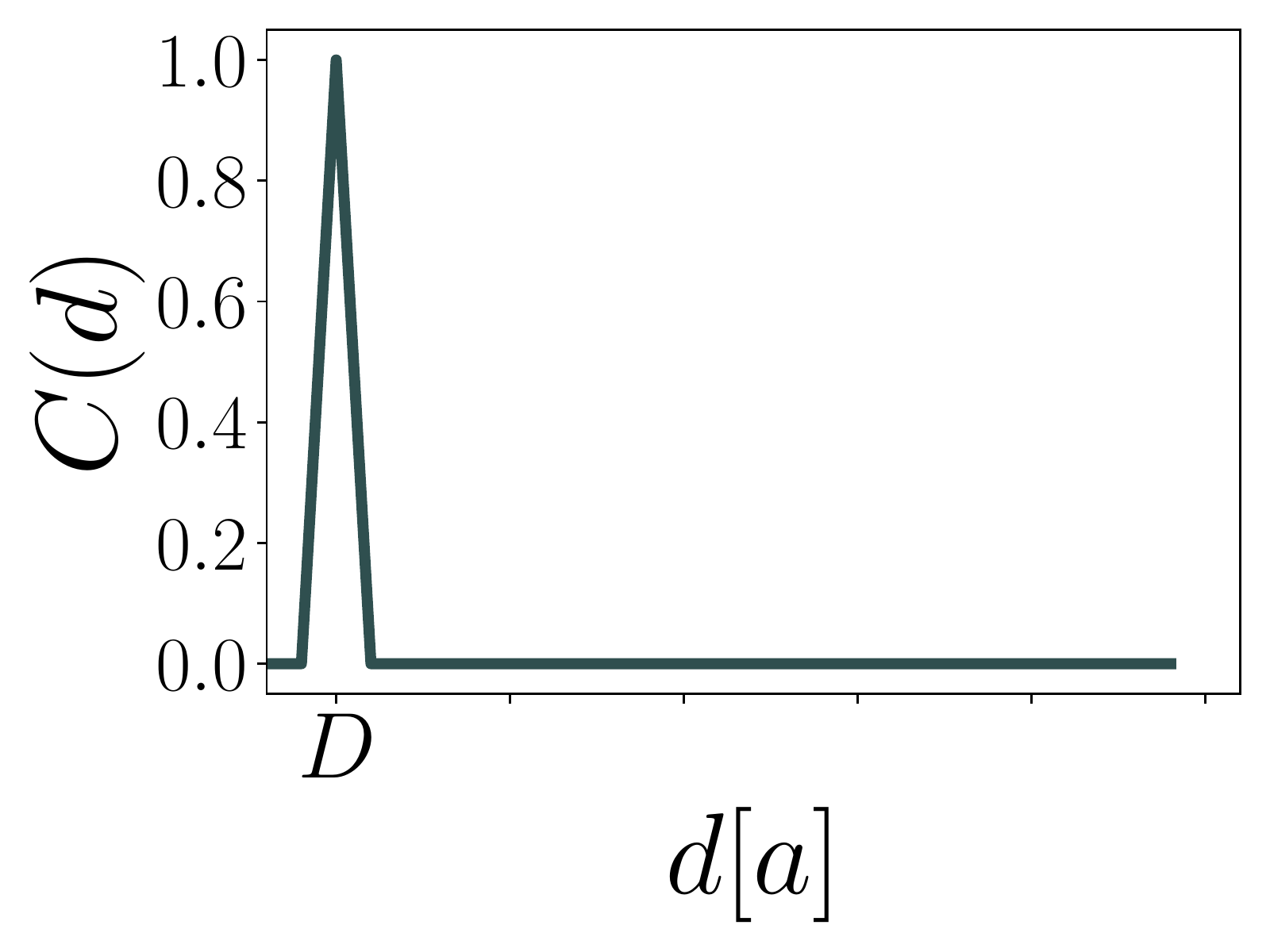} 
 \caption{ {\bf Manifestations of fracton conservation laws in antiferromagnets.} {\bf a)}: The gravitational behavior of fractons manifests as phase separation of holes.  {\bf b)}: Conservation of dipole moment $D$ manifests as a delta function in the pair correlation function: $C(d) \sim \delta(d-D)$. $d$ is in the units of the lattice constant $a$.
 }
 \label{fig:Manif}
 \end{figure}

As a consequence, holes doped into 2D and mixed-dimensional AFMs phase separate at finite hole concentrations \cite{emery,kivelson,Marder,batista}, reflective of the gravitational force between fractons.  To see this, we note that in the mixed-dimensional limit,  the model can be mapped onto a fermionic model by mapping a pair onto a spinless fermion and a spin onto an empty site \cite{batista}.  The result is $H = -t_{2} \sum_i (f_i^\dagger f_{i+1} + h.c.) - \frac{J}{4} \sum_i n_i n_{i+1}$, and $i$ runs over sites of a lattice of reduced size that results from the mapping. Here $t_2$ is an effective nearest-neighbor pair ($f$ particle) hopping that accounts for most of the pair's kinetic energy, neglecting beyond-nearest-neighbor hopping. This Hamiltonian demonstrates competition between the hopping of bound pairs ($t_2$ term) and their interaction ($J/4$ term), which is attractive due to the antiferromagnetic correlations in the background that favor spin clusters of larger size so as to increase the antiferromagnetic energy.  Thus, holes favor clustering together.  \btext{Importantly, the fractonic $t_2$ pair hopping alone is sufficient to induce an emulsion of dipolar pairs and single fractons (unpaired holes) at finite fracton `charge' densities (magnetizations), see Figure~\ref{fig:Manif}, {\bf a)}.}

\section*{Discussion}

\noindent{\em Symmetry-protected fracton order:} Topological fracton order found in special three-dimensional spin models such as the Chamon model \cite{chamon}, X-cube model \cite{fracton1} and Haah's code \cite{haah} persists against arbitrary local perturbations that are small compared to the gap regardless of symmetry considerations.  In contrast, our fracton model displays physics robust to arbitrary {\it symmetry-preserving} perturbations small in the scale of the gap, as we detail below.  This relationship parallels the one that exists between topological order and symmetry-protected topological phases; the former being robust against any perturbation while the latter survives only symmetry-respecting perturbations. We can therefore regard our model of fractons in the hole-doped Ising AFM as an example of symmetry-protected fracton (SPF) order \cite{potter,supersolid,cheng}.  Specifically, any perturbation that preserves $S_z$ conservation will still yield fractons in the hole-doped AFM, while symmetry-breaking terms would promote free hole motion, destroying fractonicity. Essentially, the ordinary global ${\rm U(1)}$ symmetry of $S_z$ conservation on the microscopic spins imposes dipole conservation $D$ on the emergent fractons. 

The idea of SPF order might serve as a guiding principle to relate symmetries to fracton phenomenology.  For example, arbitrary ${\rm U(1)}$ symmetry-preserving terms added to the Hamiltonian would not violate dipole conservation, which in turn dominates the characteristics of the system including its restricted dynamics and the emulsion fracton physics at finite hole concentrations.  In essence, this might allow us to investigate systems previously unexplored, and deduce their phenomenology on the basis of symmetry.  In the current work, the ${\rm U(1)}$ symmetry implies single hole localization, dipolar pairs and phase separation, all of which are understood as a result of emergence of fractons protected by the underlying  ${\rm U(1)}$ symmetry.

\noindent{\em Experimental relevance:} Antiferromagnetic materials \cite{Mater1} including dysprosium phosphates \cite{Mater3a, Mater3b}, dysprosium aluminum garnets \cite{Mater2}, rubidium cobalt fluorides \cite{Mater4b,Mater4a,Mater4c,Mater4d}, the quasi-one-dimensional $\kappa$-type organic salts \cite{Mater5a,Mater5b} and rare-earth pyrogermanates \cite{Mater6} serve as realistic solid-state setups to realize fractons upon doping. Rydberg-atom arrays \cite{Labuhn,Zeiher1,Zeiher2}, trapped ions \cite{Porras, Britton}, polar molecules \cite{Barnett,Gorshkov}, and ultracold atoms in optical lattices \cite{Boll, Cheuk,Mazurenko} present alternative avenues to simulate doped Ising AFMs. In two (and three) dimensions, fracton behavior is approximate.  An external field can be utilized to implement the mixed-dimensional limit for which fracton behavior becomes exact. Here, the potential gradient $V>>t$ manifests as an energetically high barrier for hole tunneling in the perpendicular direction, and remains robust on timescales $\sim (t^2/V)^{-1}$. Nano/optical photodoping techniques applied to antiferromagnetic Mott insulators \cite{Werner1} serve to engineer fractons and dipoles with long lifetimes \cite{DemlerPhoto, Werner2}, because the antiferromagnetic background functions perfectly to absorb the excess kinetic energy of the photodoped carriers on timescales of the order of few electronic hops \cite{UFPhotodoping1, UFPhotodoping2}.

We now discuss the stability of fracton behavior to perturbations beyond the Ising limit, namely a Heisenberg exchange $J_{\perp}$.  While the first application of the frustration-inducing motion of the hole creates a spin misalignment in the background (a domain wall where two neighboring spins align in the same direction), nearest-neighbor spin exchange can lift this misalignment {\em only} after a subsequent hop of the hole.  This is because after the second hop the second displaced spin becomes nearest neighbor to the oppositely oriented first one, and only then the two can flip flop, healing the background. Thus, to leading order in $t/J$ the emergent fractonicity remains stable against weak and possibly moderate $J_{\perp}$, and deviations shall occur on timescales $\sim (t^4/J^3)^{-1}$.

Interferometric and spectroscopic studies of lightly doped N\'eel AFMs serve as probes of fractons and dipoles. Absorption spectra and pair correlation functions together represent measurements that elucidate clear signatures of fractonic behavior:
Since a fracton has no dispersion, the fracton peak, the lowest pole, in the one-body spectral function ${\cal A}(k,\omega) = -\frac{1}{\pi} G_{1h}(k,\omega)$ will exhibit no dependence on $k$; A sharper diagnostic is to probe the distance between the fractons constituting a dipole: The perfect locking of the two particles within a dipole will manifest in the real-space magnon-integrated density-density correlation function: $C(d) = {\rm Tr}_{\rm magnons}  \big\langle \frac{1}{N} \sum_i \hat{n}_i\hat{n}_{i+d}\big\rangle.$
Here, $N$ is the number of lattice sites. Since, for any given two-particle state, the particles are separated by a constant distance $D$ (the dipole moment), this correlation function will be nonzero only for $d=D$, i.e. $C(d) \sim \delta(d-D)$, as shown in Figure~\ref{fig:Manif}, {\bf b)}. In contrast, a two-particle state in a system without dipole conservation would feature a more generic distribution of this correlation function, without such a sharp peak.  Note that for contexts in which fracton behavior is approximate, the density-density correlation function will feature a rounded, yet still prominent, peak in $C(d)$ near $d=1$. We wish to note that experiments studying magnetic polarons already show indications of the fracton phenomenon, including their restricted mobility and the string-mediated binding of dipoles \cite{Exp1, Exp2, Exp3, Exp4, Exp5}.  

Identifying fractons in antiferromagnets paves the way to observing their peculiar properties in transport. It was recently realized that idealized fracton models exhibit anomalous non-thermalizing behavior despite the absence of quenched disorder \cite{spread}.  For certain initial conditions, the system fails to thermalize even at asymptotically long times, analogous to the behavior of quantum many-body scars \cite{spread2,turner}.  Our analysis suggests that this behavior might emerge in hole-doped antiferromagnets for initial states respecting fractonic conservation. This possibility is supported by numerics of fractonic models \cite{pollmann, premNum}.  A versatile experimental platform like the one we propose might allow to probe exotic behavior such as the unusual late-time oscillations in certain operator quantities, speculated to occur in fracton systems as a consequence of their connection to quantum many-body scars \cite{spread2}. Fractons in antiferromagnets would also allow the exploration of unusual many-fracton phases of matter, with properties qualitatively different from usual electronic phases \cite{phases}, such as fracton microemulsions composed of small-scale clusters emulsed in a phase dominated by long-range repulsion.

\noindent{\em Further remarks:} We have identified a concrete physical realization of fractons in hole-doped Ising antiferromagnets.  \rtextnpj{While we have focused throughout on the example of the square lattice, our results apply to all bipartite lattices.} The concept of distortion-controlled motion of particles discussed here arises in various contexts and may lead to fracton behavior in matter-gauge field \cite{confine} and electron-phonon \cite{SousBerciu,SousRep} coupled systems.

Coulomb repulsion modeled as an effective $V$ plays an important role in materials.  We expect fracton behavior to survive: $V$ simply shifts the nearest-neighbor pair's energy $E_{\rm  BP}$ to $E_{\rm BP} + V$, which -- in a completely isolated system -- will be infinitely stable, since there exists no mechansim to couple this state to two free holes. At higher temperatures above the hopping scale, i.e. in the classical regime, this model exhibits a spatially heterogeneous glass phase with regions of high and low mobility, and with characteristics reminiscent of structural glasses \cite{chamon2}. Thus, this model in presence of $V$ gives rise to constrained dynamics in the quantum limit possibly leading to non-thermal states, and slow glassy dynamics in the classical limit, opening a door to investigating classical-quantum crossover in non-ergodic phenomena.

Looking ahead, one-dimensional pair-hopping models related to Equation \eqref{FracH} host topological edge modes with an unusual {\em gapless} bulk \cite{Ruhman}. Understanding the topological character of our constrained fractonic pair-hopping model may serve to expose connections between topological and fractonic behavior. \rtextnpj{Such a task might allow understanding of Haldane edge modes in hole-doped antiferromagnets \cite{GenJW} in light of fracton symmetries.} Extensions to doped two-dimensional {\em frustrated} antiferromagnets may realize more exotic types of fractons with mobility constraints extended to a line or a plane. Our work sets the stage to explore these questions.     \newline

\noindent{\bf \large Methods} \\
\noindent {\bf Self-consistent approach to the propagators.} 
The equation of motion for the one-hole propagator $G_{1h}(k,\omega) = \bra{\Psi_{\rm GS}}h_{k} \hat{G}(\omega)h^\dagger_{k}\ket{\Psi_{\rm GS}}$ and for the two-hole propagator (in the $\sigma^z_{\text{Total}} = 0$ sector) $G_{2h}(K,\omega) = \bra{\Psi_{\rm GS}}h_{k_{\sigma}} h_{k_{-\sigma}} \hat{G}(\omega)h^\dagger_{k_{-\sigma}} h^\dagger_{k_{\sigma}} \ket{\Psi_{\rm GS}}$  (here $K=k_{\sigma}+k_{-\sigma}$) in the mixed-dimensional antiferromagnet are exactly solvable in the self-consistent non-crossing scheme.  Here, $ \hat{G}(\omega) = [\omega - H]^{-1}$, $\ket{\Psi_{\rm GS}}$ represents the Ising AFM, and $H$ is in the mixed-dimensional limit. Since one hole of spin $\sigma$ always first emits a string of magnons before it absorbs them on return its original site or before they are absorbed by the second hole of spin $-\sigma$, all crossed boson lines vanish, making the non-crossing scheme  exact. A discrete pole in $G_{1h}(k,\omega)$ and similarly in $G_{2h}(K,\omega)$ signals the formation of a bound state: A magnetic polaron of dispersion $E_{\rm p}(k)$ in the one-hole case, and a magnetic bipolaron  with a dispersion $E_{\rm BP}(K)$ in the two-hole case.  See the Supplementary Note 1 for more details.

\section*{Data Availability}
The authors can confirm that all relevant data are included in the paper.

\begin{acknowledgements}
We acknowledge insightful discussions with Mona Berciu. This material is based upon work supported by the National Science Foundation (NSF) Materials Research Science and Engineering Centers (MRSEC) program through Columbia University in the Center for Precision Assembly of Superstratic and Superatomic Solids under Grant No. DMR-1420634 (J.~S.) and the Air Force Office of Scientific Research under award number FA9550-17-1-0183 (M.~P.). 
\end{acknowledgements}

\section*{Author Contributions}
Both authors contributed to the development of the ideas in this work and to the writing of the paper.

\section*{Competing Interests}
The authors declare no competing financial or non-financial interests.\newline \newline \newline

\clearpage

\onecolumngrid
\appendix

\section*{Supplementary Information}

\subsection*{Supplementary Note 1}
\noindent We compute the one- and two-hole propagators approximately for the case of two-dimensional antiferromagnets and exactly for the mixed-dimensional case.  Here we provide a brief overview of the approach in the mixed-dimensional case.

\noindent{\em One-hole propagator:} The equation of motion for the one-hole propagator $G_{1h}(k,\omega) = \bra{\Psi_{\rm GS}}h_{k} \hat{G}(\omega)h^\dagger_{k}\ket{\Psi_{\rm GS}}$ is exactly solvable in the self-consistent non-crossing scheme since crossing diagrams identically vanish as the hole must always retrace its path back to the origin.  Here, $ \hat{G}(\omega) = [\omega - H]^{-1}$, $\ket{\Psi_{\rm GS}}$ represents the Ising AFM, and $H$ is in the mixed-dimensional limit.  One then finds 
\begin{eqnarray}
G_{1h}(k,\omega) = \Big\{\left[G^0_{1h}(k,\omega)\right]^{-1}-\Sigma(k,\omega)\Big\}^{-1},
\end{eqnarray}
where $G^0_{1h}(k,\omega) = \omega^{-1}$ is the static hole propagator governed by $H_0\equiv H_{t=0}$ and the self-energy
\begin{eqnarray} 
&& \Sigma(k,\omega) = 2 t^2 F_{1h,1b}(k,\omega-3J/2)\text{,} \nonumber \\
&\text{with } F_{1h,1b}&(k,\omega-3J/2) =\Big\{\left[G^0_{1h}(k,\omega)\right]^{-1}-\Sigma_F(k,\omega)\Big\}^{-1}
\end{eqnarray}
is a generalized hole--one-boson propagator that is solved self-consistently:
\begin{eqnarray} 
\Sigma_F(k,\omega) = t^2 F_{1h,1b}(k,\omega-J).
\end{eqnarray} 
Note the different $t^2$ factors and shifts in $\omega$ in the definitions of the self-energies. These reflect the following. The first magnon created by the hole has an energy $3J/2$, corresponding to the breaking of three antiferromagnetic bonds, and can be generated in two different directions along $x$. However, subsequent magnons each cost an energy $J$, corresponding to breaking two bonds, and can only be generated unidirectionally. As such, all processes involving more than a single magnon are summed up to obtain $F_{1h,1b}$, from which one finds $G_{1h}(k,\omega)$.  This approach gives for the lowest pole $G_{1h}(k,\omega) \sim [\omega - E_{\rm p}(k)]^{-1}$, where $E_{\rm p}(k)$ is the energy dispersion of a magnetic polaron formed of the hole dressed by magnons.  
 
\noindent{\em Two-hole propagator:} The two-hole propagator in the $\sigma^z_{\text{Total}} = 0$ sector $G_{2h}(K,\omega) = \bra{\Psi_{\rm GS}}h_{k_{\sigma}} h_{k_{-\sigma}} \hat{G}(\omega)h^\dagger_{k_{-\sigma}} h^\dagger_{k_{\sigma}} \ket{\Psi_{\rm GS}}$ is also found exactly via a non-crossing self-consistent approach.  Here, $K=k_{\sigma}+k_{-\sigma}$ is the total momentum of the pair.  To see the exactness of the non-crossing scheme for two-hole states, note that one hole of spin $\sigma$ always first emits a string of magnons before they are absorbed by the second hole of spin $-\sigma$; this represents the only possible magnon-mediated interaction between the holes, and thus all crossed boson lines connecting the two fermion lines vanish.   One then sums all diagrams self-consistently (all crossed boson lines vanish) to find the ${\mathcal T}$-matrix describing the interaction between two holes. A discrete pole in $G_{2h}(K,\omega)$ signals the formation of a bound state of two holes, with a dispersion $E_{\rm BP}(K)$.   Note that such a bound state has an infinite lifetime even if it is not the lowest energy state, since there are no matrix elements to couple the bound pair to a configuration of two isolated particles.

\end{document}